\documentstyle[a4,12pt,epsf]{article}
\newcommand{\noi}{\noindent}
\newcommand{\eq}{\begin{equation}}
\newcommand{\en}{\end{equation}}
\newcommand{\eqa}{\begin{eqnarray}}
\newcommand{\ena}{\end{eqnarray}}

\newcommand{\vx}{{\vec x}}
\newcommand{\vxp}{{\vec x}_{\perp}}
\newcommand{\vp}{{\vec p}}

\newcommand{\vq}{{\vec q}}
\newcommand{\vqp}{{\vec q}_{\perp}}
\newcommand{\vR}{{\vec R}}

\newcommand{\bpartial}{{\bar \partial}}
\newcommand{\eff}{e\! f\! f}

\newcommand{\cak}{{\cal K}}
\newcommand{\cao}{{\cal O}}

\newcommand{\lra}{\longrightarrow}

\newcommand{\meff}{m_{e\! f\! f}}
\begin{document}

\renewcommand{\baselinestretch}{1.1}
\small\normalsize
\renewcommand{\theequation}{\arabic{section}.\arabic{equation}}
\renewcommand{\thesection}{\arabic{section}}
\language0

\vspace{-2cm}

\hbox{}
\noindent  March 1999 \hfill   JINR E2--99--43\par
                             \hfill   SHEP/99/01

\vspace{0.5cm}
\begin{center}

\renewcommand{\thefootnote}{\fnsymbol{footnote}}
\setcounter{footnote}{1}

{\LARGE Gribov copies and gauge variant correlators in $U(1)$ lattice 
gauge theory} 

\end{center}

\vspace*{0.5cm}
{\large
I. L. Bogolubsky$\mbox{}^1$,
L. Del Debbio$\mbox{}^2$ and
V.K.~Mitrjushkin$\mbox{}^1$ 
}\\

\vspace{0.5cm}

{\normalsize
$\mbox{}^1$ {\em Joint Institute for Nuclear Research, Dubna, Russia} \par
$\mbox{}^2$ {\em Dept. of Physics and Astronomy, Univ. of 
Southampton SO17 1BJ, UK } \\
}     
\vspace*{0.5cm}

\begin{center}
{\bf Abstract} 
\end{center}

We discuss the influence of Dirac sheets and zero--momentum modes on
the gauge variant photon correlators $~\Gamma(\tau;\vp)~$ with
$~\vp\ne 0~$ and $~\vp =0~$ in the pure gauge $~U(1)~$ theory. 
A special attention has been paid to the $\beta$- and volume--dependence
of this influence.  Numerical simulations are performed on $~12\times
6^3~$ and $~24\times 12^3~$ lattices at different $~\beta$'s in the
Coulomb phase.

      
\section{Introduction} \setcounter{equation}{0}

Lattice approach \cite{wil} gives a possibility to calculate gauge
invariant objects without gauge fixing. However, numerical calculations
of gauge dependent objects, e.g., gauge field correlators
$~\Gamma(\tau;\vp)~$, can also be of interest. For example, the
comparison of gauge variant objects with that calculated perturbatively
can provide some deeper insight into the structure of the lattice theory.
In particular, the role of lattice artifacts and/or Gribov copies
\cite{grib} can be investigated.

Compact $U(1)~$ pure gauge theory in the Coulomb phase provides a 
unique `test ground' for the lattice approach. Indeed, in the weak 
coupling limit one expects to find a trivial theory of the free 
non--interacting photons.  Therefore, at sufficiently small coupling a 
gauge variant transverse correlator $~\Gamma_T(\tau;\vp)~$ is supposed 
to fit the perturbative expression given in eq.(\ref{cor_pert}).

A few years ago it has been shown \cite{nak} that in the Coulomb phase
some of the gauge copies produce a photon correlator
$~\Gamma_T(\tau;\vp\ne 0)~$ with a decay behavior inconsistent with
perturbation theory.  Numerical study \cite{bmmp} has shown that there
is a connection between `bad' gauge copies and the appearance of
configurations with periodically closed double Dirac sheets (DDS). The
explanation of this effect \cite{clas} is connected with a nontrivial
`vacuum' structure of the classical compact $~U(1)~$ theory, i.e. with
the existence of the nontrivial stable classical solutions (`vacua'),
some of these solutions -- DDS -- being Gribov copies of the trivial
`vacuum' with zero field.

Another interesting observation is connected with the
$~\tau$--dependence of the zero--momentum gauge variant correlators
$~\Gamma(\tau)~$.  The analysis of the $~\tau$--dependence of the
correlator $~\Gamma(\tau)~$ in the $~SU(3)~$ lattice gauge theory drove
the authors of \cite{mo1} to the conclusion that $~\Gamma(\tau)~$ is
consistent with the propagation of a massive gluon.  On the other side,
in the $~U(1)~$ theory the correlator $~\Gamma(\tau)~$ exhibits a
similar behaviour \cite{mo2}.
However, one can hardly expect the appearance of the massive
photon in the Coulomb phase in the $~U(1)~$ theory.
It has been shown in ref.~\cite{zmod} that (at least, in the case
of the $~U(1)~$ theory) the $~\tau$--dependence of the correlator
$~\Gamma(\tau)~$ can be explained taking into account zero--momentum
mode of the gauge field.
The influence of the zero--momentum modes on the gluon propagators in
nonabelian theories has been pointed in \cite{at1,at2,su2}.

In this paper we discuss the effects connected with Dirac sheets and
zero--momentum modes in more details. The main questions addressed in this
paper are the following.

\noi -- What is the $~\beta$--dependence of the frequency of the Dirac
sheets appearance ?  Does the contribution of the Dirac sheets to the gauge
variant correlator $~\Gamma(\tau;\vp)~$ become more (or less)
important when the coupling $~\beta~$ is increased?

\noi -- What is the volume dependence of the correlator
$~\Gamma_T(\tau;\vp)~$ ?  Does the influence of DDS become smaller in
the thermodynamical limit ?

\noi  -- What is the influence of zero--momentum modes on the
zero--momentum correlator $\Gamma(\tau)$ ? How does this
influence depend on $~\beta~$ and lattice size ?

Throughout this paper a $~4d~$ lattice 
with periodic boundary conditions is considered.
$N_{\mu}$ is the lattice size in the direction $~\mu$, and
$~V_4 =N_1 N_2 N_3 N_4$.
The lattice derivatives are
$~\partial_{\mu} f(x) = f(x+{\hat \mu}) -f(x)~$ and
$~\bpartial_{\mu} f(x) = f(x) - f(x-{\hat \mu})$,
and the lattice spacing is chosen to be unity.

\section{The action and classical solutions}
 \setcounter{equation}{0}

The standard Wilson action $~S(U)~$ is 

\eq
S(U) = \beta \sum_{x} \sum_{\mu >\nu}
        \,  \Bigl( 1 - \cos \theta_{x;\mu\nu} \Bigr) ~,
\en

\noi where $U_{x\mu} = e^{i\theta_{x\mu}} \in U(1)$ are link 
variables, $~\theta_{x;\mu\nu} = \partial_{\mu} \theta_{x\nu} 
- \partial_{\nu} \theta_{x\mu}~$ are the plaquette angles and
$~\beta = 1/g^2~$.
This action  is the part of the full QED action $S_{QED}$, which is
supposed to be compact if we consider QED as arising from a subgroup of
a non--abelian (e.g., grand unified) gauge theory \cite{pol2}.

The plaquette angle $~\theta_P \equiv \theta_{x;\, \mu \nu}~$
can be split up: $~\theta_P = [\theta_P] + 2\pi n_P$, where
$~[\theta_{P}] \in (-\pi ;\pi ]$ and $n_P = 0, \pm 1, \pm 2$.
The plaquettes with $~n_P \neq 0~$ are called Dirac plaquettes.
The dual integer valued plaquettes $~m_{x,\mu \nu} =
\frac {1}{2} \varepsilon_{\mu \nu \rho \sigma} n_{x,\rho \sigma}~$
form Dirac sheets \cite{dgt}.

In a perturbative language the existence of the gauge copies can be
interpreted as a problem of the gauge copies of the `vacuum'
configurations, i.e.  the solutions of the classical equations of
motion.  Assuming that every configuration is some small fluctuation
about the corresponding `vacuum' one can find all gauge copies of this
configuration provided all gauge copies of the `vacuum' are known.

The classical equations of motion are

\eq
\sum_{\nu} \bpartial_{\nu} \sin \theta^{cl}_{x;\mu \nu} = 0~.
                  \label{class_eq_u1}
\en

\noi Evidently, zero--momentum modes $~\theta^{cl}_{x\mu}=\phi_{\mu}~$ 
are the solutions of these equations.

Choosing the Lorentz (or Landau) gauge
$~\sum_{\mu} \bpartial_{\mu} \theta_{x\mu} =0~$,
one can write expicitely a solution corresponding to a
single Dirac sheet \cite{clas}

\eqa
\theta^{cl}_{x1}(\vR ) &=& \frac{2\pi i }{N_1N_2}
\sum_{\vqp \ne 0} \frac{\cak_2} {{\vec \cak }^{\, 2}_{\perp}}
\cdot e^{i\vqp (\vxp - \vR ) - \frac{i}{2}q_2 }~;
                 \label{sds_u1}
\\
\nonumber \\
\theta^{cl}_{x2}(\vR ) &=& -\frac{2\pi i }{N_1N_2}
\sum_{\vqp \ne 0} \frac{\cak_1} {{\vec \cak }^{\, 2}_{\perp}}
\cdot e^{i\vqp (\vxp - \vR ) - \frac{i}{2}q_1 }~,
\ena

\noi where $~\cak_{\mu} =2\sin \frac{q_{\mu}}{2}$ and
$~{\vec \cak}^2_{\perp} = \cak_1^2+\cak_2^2$. 
The two--dimensional vector $\vR =(R_1;R_2)$ corresponds to
the position of the Dirac plaquette in the $(x_1;x_2)$ plane:
$$~\theta^{cl}_{x;12} = 2\pi \cdot \delta_{\vxp ;\vR} 
-\Delta~; 
\qquad \Delta = \frac{2\pi}{N_1N_2}~.$$
Of course, $~\theta^{\prime}_{x\mu} =\phi_{\mu} + \theta^{cl}_{x\mu}~$
is also a solution of eq's.(\ref{class_eq_u1}).

The single Dirac sheet solution $\theta^{cl}_{x\mu}(\vR )$ corresponds
to a local minimum of the action, i.e. that it is stable with respect to
small fluctuations. The existence of the long--living metastable states
corresponding to single Dirac sheets was observed in simulations in the
pure gauge $U(1)$ theory \cite{neu}.

The classical gauge action is
$~S(\theta^{cl} ) =\frac{2V_4}{g^2} \Bigl( 1-\cos \Delta \Bigr)~$.
On a symmetric lattice $~N_1=N_2=N_3=N_4 \to\infty~$ the action 
$~S(\theta^{cl})~$ is non--zero and finite :

$$S(\theta^{cl} )=\frac{4\pi^2}{g^2} < \infty~.$$

Double Dirac sheet solution of the classical equation of motion consists
of the two single Dirac sheets with an opposite orientation of the flux
:

\eq
\theta^{cl}_{xi}(\vR^{a};\vR^{b} ) 
= \theta^{cl}_{xi}(\vR^{a}) - \theta^{cl}_{xi}(\vR^{b}),
\qquad i=1;2~,
                \label{dds}
\en 
 
\noi where the vectors $\vR^{a}$ and $\vR^{b}$ correspond to the two 
Dirac plaquettes in the plane $(x_1;x_2)$. It is easy to see that

\eq
\theta^{cl}_{x;12}(\vR^a;\vR^b ) =2\pi \cdot \Bigl[ \delta_{\vxp ;\vR^a}
-\delta_{\vxp ;\vR^b} \Bigr]~.
\en 

\noi The double Dirac sheet $~\theta^{cl}_{x;i}(\vR^a;\vR^b )~$ has a
zero action: $~S(\theta^{cl} ) = 0$. 

Gauge transformations can shift the Dirac sheets and change their form.
For example, the `big' gauge transformation function $~\Omega_x~$

\eq
\Omega_{x} = -\frac{2\pi }{N_1N_2} \sum_{\vqp \ne 0} 
\frac{ e^{i\vqp (\vxp - \vR) }}{{\vec \cak }^{\, 2}_{\perp} } 
\cdot \Bigl( 1 - e^{-iq_2 } \Bigr)~;
                 \label{shift_x}
\en

\noi shifts the single Dirac sheet in the $x_1$--direction.
The gauge transformation $\Omega_x$ 
in eq.(\ref{shift_x}) applied to
the zero--field creates a double Dirac sheet as  in
eq.(\ref{dds}) with $\vR^a =\vR $ and $\vR^b =\vR -{\hat 1}$ .
Therefore,  $~\theta^{cl}_{x;i}(\vR^a;\vR^b )$ 
is a Gribov copy of the zero solution $~\theta^{cl}_{x\mu} =0$.
It is not difficult now to obtain general Dirac sheet 
solutions, i.e. the Dirac sheets curved in the four-dimensional
space  \cite{clas}.

\section{$~\Gamma_T(\tau;\vp)$ and Dirac sheets}
 \setcounter{equation}{0}

In lattice calculations the usual choice of the Lorentz (or Landau)
gauge is

\eq
\sum_{\mu=1}^4 \bpartial_{\mu} \sin \theta_{x\mu} = 0,
                                                  \label{gf1}
\en

\noi which is equivalent to finding extremum of the 
functional $F(\theta)$

\eq
F(\theta) = \frac{1}{V_4} \sum_{x} F_{x}(\theta )~;
\qquad F_{x}(\theta) = \frac{1}{8}\sum_{\mu=1}^4
\Bigl[ \cos \theta_{x\mu}+ \cos \theta_{x-{\hat \mu};\mu}\Bigr]
                                                 \label{gf2}
\en

\noi with respect to gauge transformations
$~U_{x\mu} \lra \Lambda_{x} U_{x \mu} \Lambda_{x+\mu}^{\ast}~;
~~\Lambda_x=\exp\{i\Omega_x\}~$.
It is important to note that all link angles $\theta_{x\mu}$ are
compact variables  ($ -\pi < \theta_{x\mu} \le \pi $), and the
gauge transformations 
$\theta_{x\mu} \stackrel
{\Omega}{\to} \theta_{x\mu} - \partial_{\mu} \Omega_x$
 are understood modulo $2\pi$.

To find extremum of $~F(\theta)~$ we carried out a standard
overrelaxation gauge cooling procedure with parameter $~\alpha~$ tuned
to minimize the number of the cooling steps for given $~\beta~$ and
volume. In our computations values of $~\alpha~$ were chosen to be
between $~\alpha=1.66~$ and  $~\alpha=1.84~$.  The stopping criterion
was

$$ \max \left |\sum_{\mu} \bpartial_{\mu}\sin \theta_{x\mu} \right |<
10^{-5} \quad \mbox{and} \quad \frac{1}{V} \sum_x \left|\sum_{\mu}
\bpartial_{\mu}\sin \theta_{x\mu} \right |< 10^{-6}~. $$

The photon correlator $~\Gamma_{\mu}(\tau ;\vp )~$ is

\eqa
\Gamma_{\mu} (\tau ;\vp ) &=& \Bigl\langle 
{\cal O}_{\mu}^{\ast}(\tau ;\vp ) {\cal O}_{\mu}(0;\vp ) \Bigr\rangle
= \frac{1}{N_4}\sum_{t=0}^{N_4-1} \Bigl\langle 
{\cal O}_{\mu}^{\ast}(t \oplus \tau ;\vp ) {\cal O}_{\mu}(t;\vp ) \Bigr\rangle~;
\ena

\noi where $t \oplus \tau = (t+\tau )\mbox{ mod } N_4~$ and 

\eq
{\cal O}_{\mu}(\tau ;\vp ) = \sum_{\vx} e^{-i\vp \vx -\frac{i}{2}p_{\mu}} 
\cdot \sin \theta_{x\mu}~,
\qquad \mu =1,2,3~,
                    \label{oper_gau_dep}
\en

\noi and, clearly,  $~\langle O_{\mu} \rangle =0$.

Let us choose  the momentum $~\vp =(0;p_2;0)~$ with $~p_2 \ne 0$, and 
$\mu =1$.
The perturbative expansion about the zero solution of the classical
equation of motion $\theta^{cl}_{x\mu} =0$, i.e.  the standard
perturbation theory, gives in the lowest approximation

\eq
\Gamma^{pert}_1(\tau;\vp ) \sim e^{-\tau E_p }+e^{-(N_4 - \tau )E_p}~,
               \label{cor_pert}
\en

\noi where the energy $E_p$ satisfies the lattice dispersion
relation

\eq
\sinh^2 \frac{E_p}{2} = \sum_{i=1}^3 \sin^2 \frac{p_i}{2}~.
               \label{ldr}
\en

It is easy to see that the expansion about a Dirac sheet solution
gives a contribution to the correlator very different from that in
eq.({\ref{cor_pert}).  

As an example, let us choose the flat double Dirac sheet with the
space--like Dirac plaquettes $~\theta^{cl}_{x\mu}(\vR_1 ;\vR_2 )$ as
defined in eq.(\ref{dds}).  In this case the correlator in the double
Dirac sheet background is

\eq
\Gamma^{dds}_{1} (\tau ;\vp ) = \Bigl| \Phi (\vp )\Bigr|^2
+ \frac{g^2}{2V_3} \sum_{\vqp} G(\tau ;\vqp + \vp ) 
\cdot \Bigl| \Psi(\vqp ) \Bigr|^2~,
               \label{cor_3}
\en

\noi where

\eq
\Phi (\vq ) = \frac{\delta_{q_3;0}}{N_1N_2}  
\sum_{\vxp} e^{-i\vqp \vxp } \sin \theta^{cl}_{x1} ~;
\quad
\Psi (\vq ) = \frac{\delta_{q_3;0}}{N_1N_2}  
\sum_{\vxp} e^{-i\vqp \vxp } \cos \theta^{cl}_{x1} ~,
\nonumber 
\en

\noi and

\eqa
G(\tau ;\vq ) &=& \left[ 1 + \frac{\cak_1^2(q)}{2\sinh E_q} 
\cdot \frac{d}{dE_q } \right] G_0(\tau ; \vq )~;
\\
\nonumber \\
G_0(\tau;\vq ) &=& \frac{1}{2 \sinh E_q} 
\frac{1}{1 - e^{-N_4E_q}} \cdot
\Bigl[ e^{-\tau E_q } + e^{-(N_4 - \tau )E_q} \Bigr] +\ldots ~.
\nonumber
\ena

This expression is obviously different from that given in
eq.(\ref{cor_pert}). In particular, the first term in the r.h.s.
in eq.(\ref{cor_3}) produces the `shift upward' of the correlator
with respect to the perturbative expression.

In our computations we monitored the total number of the 
Dirac plaquettes $~N^{(\mu\nu)}_{DP}~$ for every plane $~(\mu\nu)~$ and 
$~N_{DP}= \max_{(\mu\nu)} N^{(\mu\nu)}_{DP}~$.  By a detailed 
investigation of the number and the location of Dirac plaquettes in the 
Coulomb phase, we observed pairs of Dirac sheets closed by periodic 
boundary conditions. In the extreme case of a minimal surface this is a 
pair of periodic Dirac sheets occupying parallel planes and having 
corresponding Dirac plaquettes with opposite signs.  The appearance of 
periodically closed double Dirac sheet means that the number of the
Dirac plaquettes in a given 
plane $~(\mu\nu)~$ has to be 
$$~N^{(\mu\nu)}_{DP} \ge \frac{2V_4}{N_{\mu}N_{\nu}}~.$$ 

For example, on a $~12\times 6^3~$ lattice, the minimal number of Dirac 
plaquettes is $~N_{DP}^{min}=72~$ if the Dirac sheets are timelike (`short' 
DDS), and $~N^{min}_{DP}=144~$ if the Dirac sheets are spacelike (`long' 
DDS).

Figure \ref{fig:ndpm_2lat_2beta_LG}a shows a time history of the number
of the Dirac plaquettes $~N_{DP}~$ on the $~12\times 6^3~$ lattice at
$~\beta =1.1~$. For every configuration the number $~N_{DP}~$ has been
measured after the gauge fixing procedure.  The number of configurations
with Dirac sheets ($N_{DP} \ge 72$) is $~\sim 20\% ~$.  Long
DDS ($N_{DP} \ge 144$) are not as frequent as short DDS. However, as we
shall see later the influence of the long DDS on the photon correlator
is much stronger.

To convince ourselves that the appearance of the Dirac sheets is not an 
artifact of the small volume and/or small $~\beta$--values we repeated 
these calculations on the lattices $~12\times 6^3~$ and $~24\times 
12^3~$ at $~\beta$--values between $~\beta=1.1~$ and $~\beta=10~$.  We 
observed that the probability to find DDS shows rather weak 
dependence on the coupling. We observed also that the probability to find
DDS increases slightly with increasing of the volume (e.g., at $~\beta=2~$
the number of configurations with DDS is $~\sim 17\%~$ on the 
$~12\times 6^3~$ lattice and $~\sim 25\%~$ on the 
$~24\times 12^3~$ lattice).
In Figure \ref{fig:ndpm_2lat_2beta_LG}b we show
a time history of the number of the Dirac plaquettes $~N_{DP}~$ on the 
$~24\times 12^3~$ lattice at $~\beta =2~$. 
This time history looks less `noisy' as compared with that in 
Figure \ref{fig:ndpm_2lat_2beta_LG}a
because there are less monopoles at $~\beta=2~$.

In Figure \ref{fig:pcr_2lat_2beta_LG}a
we show the transverse correlator $~\Gamma(\tau;{\vec p})~$ with 
$~{\vec p} = (0;\frac{2\pi}{N_2};0)~$ at $~\beta=1.1~$ on the 
$~12\times 6^3~$ lattice.  
Solid line corresponds to the lowest order perturbative expression 
given in eq.(\ref{cor_pert}). Averaging over all configurations (filled
circles) gives noticeable disagreement with the free photon case.  On 
the other hand, excluding from the statistics configurations containing 
Dirac sheets gives an average correlator (opened circles) which fits 
nicely the zero--mass photon correlator. Squares represent
a correlator obtained by averaging over configurations containing 
long Dirac sheets.
The destructive influence of the Dirac sheets exhibits little changes
with increasing $~\beta~$ and the lattice size.
As an example we show in Figure \ref{fig:pcr_2lat_2beta_LG}b
the transverse correlator $~\Gamma(\tau;{\vec p})~$ 
at $~\beta=2~$ on the $~24\times 12^3~$ lattice.  

The results presented in this section could be summarised by saying
that the contribution of configurations with DDS `spoils' the photon
correlator and leads to a wrong dispersion relation inconsistent with
the dispersion relation for the massless photon. Dirac sheets
represent an example of lattice artifacts which can lead to a
wrong interpretation of the results of numerical calculations.

\section{Zero--momentum correlator}
 \setcounter{equation}{0}

The (connected) zero--momentum correlator $~\Gamma(\tau)~$ is

\eq
\Gamma_{\mu}(\tau ) = \left\langle 
{\cal O}_{\mu}(\tau ){\cal O}_{\mu}(0) \right\rangle;
\qquad
{\cal O}_{\mu}(\tau ) = \sum_{\vx} \sin \theta_{x\mu}~,
                    \label{cor_zeromom}
\en

\noi where $~\tau~$ is the separation in one of the four euclidian
directions and $~\vx~$ corresponds to the three complementary
directions.

A usual way to define effective masses $\meff(\tau )$ is

\eq
\frac{\cosh \left[\meff (\tau ) (\tau +1-\frac{1}{2}N_4)\right]}
{\cosh \left[\meff (\tau ) (\tau -\frac{1}{2}N_4)\right]}
= \frac{\Gamma_{\mu}(\tau +1)}{\Gamma_{\mu}(\tau )}~.
              \label{m_eff}
\en

\noi In the $SU(3)$ theory the large--$\tau$ behaviour of $\meff(\tau )$
is supposed to describe the `gluon mass' $m_g$ if $\tau$ is chosen along
the `temperature' direction $x_4$ and $~\mu=1,2,3~$ \cite{mo1}.

The definition of the effective mass in eq.(\ref{m_eff}) is such that
for a correlator behaving like $~\Gamma_i(\tau) \sim
e^{-M\tau}+e^{-M(N_4-\tau)}~$, one
obtains $~\meff=M~$. However, on the periodic torus the gauge variant
correlator $~\Gamma_i(\tau)~$ cannot be represented as a sum of
exponents. Zero--momentum modes make their contribution to this 
correlator adding a nonzero positive constant as in eq.(\ref{cor_zm})
\cite{zmod}. As we shall see later this contribution cannot be
discarded.

First, it is rather instructive to look at the distributions of the
operators $~\cao^{(k)}(\tau)~$ defined in eq.(\ref{cor_zeromom}) and
the correlations between different time--slices $\tau$, where the index
$k$ stands for the $k^{th}$ measurement.
The width of the distribution of the zero--momentum operators is not
small, and shows very weak dependence on the coupling constant in the
interval between $\beta =1.1$ and $\beta = 10.0$.  
As an illustration we show in Figure \ref{fig:op12_24x12_b02p00}
a `scatter--plot' of the operators $~\cao^{(k)}(\tau)~$
with $\tau =1$ and $\tau =2$ at $~\beta=2~$ on a
$~24\times 12^3~$ lattice. 
For every measurement $k$ the operators $~\cao^{(k)}(\tau)~$
can be represented as
\eq
\cao^{(k)}(\tau) = C^{(k)} + \delta \cao^{(k)}(\tau)~,
\en

\noi where $~\delta \cao^{(k)}(\tau)~$ are small fluctuations about 
the constant contribution $~C^{(k)}~$.
The width of the stripe in Figure \ref{fig:op12_24x12_b02p00}
reflects the amplitude of these fluctuations.

 It is worthwhile to note that correlations between operators
$~\cao^{(k)}(\tau;\vp)~$ with nonzero momentum are very different from
that for $~\cao^{(k)}(\tau)~$, namely the scatter plot for
$~\cao^{(k)}(\tau;\vp)~$ consists of a small `blob' at the center of
coordinates (is not shown).

It is not difficult to calculate the gauge variant correlator in the
lowest approximation.
The average of any functional $\Phi (\theta )$ is defined as

\eq
\langle \Phi \rangle = \frac{1}{Z} \int
\! [d\theta_{x\mu}] ~ \Phi (\theta ) \cdot e^{- S_{\eff}(\theta )}~,
                        \label{func_phi}
\en

\noi where $~S_{\eff}=S+S_{gf}+S_{FP}~$,
$~S_{gf} =  \frac{1}{\alpha g^2} \sum_x 
\Bigl( \sum_{\mu} \bpartial_{\mu} \sin \theta_{x\mu} \Bigr)^2$,
and $S_{FP}$ is the contribution from the corresponding
Faddeev--Popov determinant.
The choice $~\alpha =0~$ corresponds to the Lorentz gauge.

Lorentz gauge condition, does not exclude the appearance of the
zero--momentum modes in the periodic volume. To calculate $\langle \Phi
\rangle$ perturbatively one should keep the zero--momentum mode under
control in the perturbation expansion.  This can be achieved by
repeating the Faddeev--Popov trick

\eq
1 = J_0\int_{-\pi}^{\pi} \! \prod_{\mu =1}^{4} \! d\phi_{\mu}~
\exp \left\{-\frac{1}{\epsilon}\sum_{\mu} 
\Bigl(\phi_{\mu} - \frac{1}{V_4} \sum_x\theta_{x\mu} \Bigr)^2 
\right\} \Big|_{\epsilon \to 0}~,
\en

\noi and making the change of variables $~\theta_{x\mu} = \phi_{\mu}
+gA_{x\mu}$.  The average functional $\langle \Phi \rangle$ can be now
represented in the form

\eq
\langle \Phi \rangle \sim \int_{-\pi}^{\pi} \!
[d\phi_{\mu}] \! \int \! [dA_{x \mu}] 
\prod_{\mu=1}^4 \! \delta \Bigl( \sum_x A_{x\mu} \Bigr) 
\Phi (\phi +gA) e^{- S_{\eff}(\phi +gA)}~.
                        \label{func_phi_new}
\en

\noi In eq.~(\ref{func_phi_new}) one can safely expand in powers of
$g^2$ and extend the limits of integration of $A_{x\mu}$ to
$[-\infty ;\infty ]$.
The zero--momentum modes are not gaussian (the gauge action $S(\theta )$
does not depend on $\phi_{\mu}$), and the 
integration over $\phi_{\mu}$ in eq.(\ref{func_phi_new})
should stay compact.

Applying a perturbative expansion about the constant
(zero--momentum) modes $\phi_{\mu}$ to calculate 
the zero--momentum correlators $\Gamma_{\mu} (\tau )$ one obtains

\eq
\Gamma_{\mu} (\tau ) 
\sim \langle \sin^2 \phi_{\mu} \rangle \cdot (1-b_{\mu}g^2)
+ \frac{g^2}{2V_4} \langle \cos^2 \phi_{\mu} \rangle \cdot
\Gamma^{\prime}_{\mu} (\tau)~,
                 \label{cor_zm}
\en

\noi where 

\eqa
\Gamma^{\prime}_i (\tau ) &=& \frac{g^2V_3^2}{2V_4}
\sum_{p_4 \ne 0} \frac{e^{i\tau p_4}}{4\sin^2\frac{p_4}{2}}~;
\qquad i=1;2;3~;
                 \label{cor_nozm_u1}
\\
\Gamma^{\prime}_4 (\tau ) &=&0  \quad \mbox{at} \quad \alpha =0,
\nonumber
\ena

\noi and $V_3=N_1N_2N_3$.
Coefficients $b_{\mu}$ represent tadpole contributions :

\eq
b_{\mu} = \frac{1}{2V_4} \sum_{p\ne 0} \frac{1}{\cak^2}
\left[ 1- \frac{\cak^2_{\mu}}{\cak^2}\right]~;
\qquad \cak_{\mu}=2\sin\frac{p_{\mu}}{2}~.
\en

\noi The coefficients
$~\langle \sin^2 \phi_{\mu} \rangle~$ and 
$~\langle \cos^2 \phi_{\mu} \rangle~$ have been numerically calculated 
from distributions of the zero--momentum operators.

As an example we show in Figure \ref{fig:pcr_mom0_12x06_b02p00_LG} the 
correlator $~\Gamma(\tau)\equiv \Gamma_i(\tau)~$ at $~\beta=2~$ on a 
$~12\times 6^3~$ lattice. Averaging over configurations without DDS 
(filled circles) gives a nice agreement with correlator defined in 
eq.(\ref{cor_zm})  (solid line).  The agreement becomes even better 
with increasing $~\beta~$. The increasing of the lattice size does 
change this conclusion.  It is worthwhile to stress that this agreement 
is impossible without taking into account zero--momentum modes.  
Correlators defined on the configurations with DDS (diamonds and 
triangles) are very different from that defined on the configurations 
with DDS excluded.

\section{Conclusions} \setcounter{equation}{0}

We have studied the influence of Dirac sheets and zero--momentum
modes on the gauge variant correlators $~\Gamma(\tau;\vp)~$ 
with zero and nonzero momenta.
All calculations have been performed on $~12\times 6^3~$ and $~24\times 
12^3~$ lattices at different $~\beta$'s in the Coulomb phase.

We confirm our previous conclusion \cite{bmmp,clas} about the role of
the Dirac sheets : gauge copies which possess double Dirac sheets
contribute to a wrong (`nonphysical') behaviour of the gauge variant
photon propagators $~\Gamma(\tau;\vp)~$ at $~\vp =0~$ and $~\vp \ne 
0~$. 

We have checked that with the increasing of the coupling from 
$~\beta=1.1~$ up to $~\beta=10~$ on a $~12\times 6^3~$ lattice 
and from $~\beta=1.1~$ up to $~\beta=2~$ on a $~24\times 12^3~$ 
lattice the frequency of DDS does not change significantly.
This frequency grows somewhat with the increasing of the volume.
Because of configurations with DDS the numerically measured 
gauge variant correlators do not correspond to the free massless 
photons. Therefore, to avoid nonphysical conclusions one should exclude
configurations with Dirac sheets out of consideration.

The contribution of the zero--momentum modes is of crucial importance 
for the interpretation of the zero--momentum correlators 
$~\Gamma_{\mu}(\tau)~$.  These modes cannot be discarded at all 
$~\beta$'s and volumes we employed. A formal application of the 
definition of the effective mass $~\meff~$ given in eq.(\ref{m_eff}) 
can produce a misleading interpretation.

\section*{Acknowledgements}

One of us (VKM) gratefully acknowledge support obtained from
the grant INTAS 96--0370 and the JINR Heisenberg--Landau grant.
LDD is supported by PPARC under grants GR/L56329 and GR/L29927.
Two of us (VKM and LDD) are pleased to thank the Physics Department of
the University of Wales Swansea, where this work begun, for providing
a stimulating working environment.


\par

\vspace{0.5cm}
\section*{Figure captions}

\noi {Figure \ref{fig:ndpm_2lat_2beta_LG}.
~~Time history of $N_{DP}$ at $\beta=1.1$ on the $12\times 6^3$
lattice ({\bf a}) and at $\beta=2$ on the $24\times 12^3$ lattice
({\bf b}).

\vspace{0.25cm}
\noi {Figure \ref{fig:pcr_2lat_2beta_LG}.
~~$\Gamma_T(\tau;\vp )$ at $\beta=1.1$ on the $12\times 6^3$ lattice
({\bf a}) and
at $\beta=2$ on the $24\times 12^3$ lattice ({\bf b}).

\vspace{0.25cm}
\noi {Figure \ref{fig:op12_24x12_b02p00}.
~~Scatter--plot for operators $\cao(\tau =1)$
and $\cao(\tau =2)$ at $\beta=2$ on the $24\times 12^3$ lattice.

\vspace{0.25cm}
\noi {Figure \ref{fig:pcr_mom0_12x06_b02p00_LG}.
~~$\Gamma(\tau)$ at $\beta=2$ on the $12\times6^3$

\newpage


%
%
\begin{figure}[pt]
\begin{center}
\vskip -1.5truecm
\leavevmode
\hbox{
\epsfysize=14cm
\epsfxsize=14cm
\epsfbox{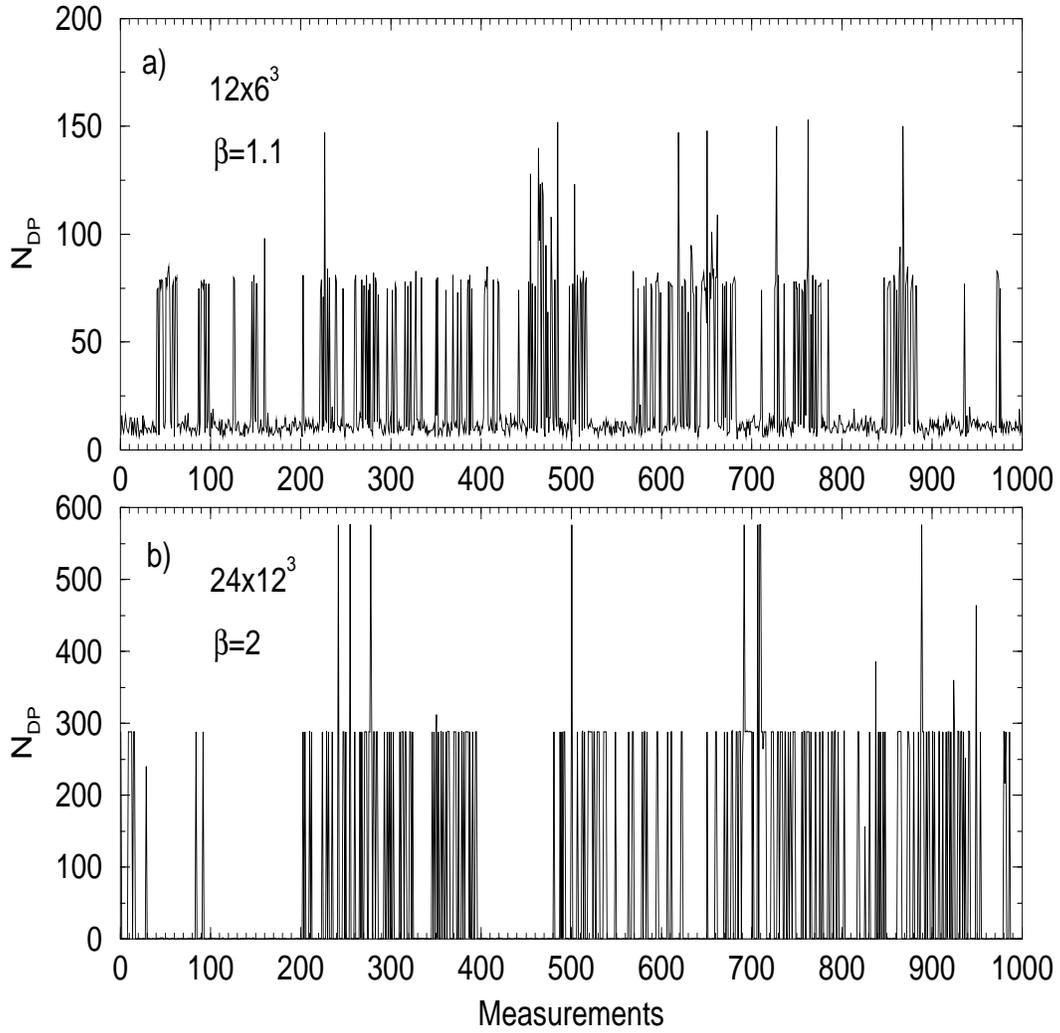}
}
\end{center}
\caption{Time history of $N_{DP}$ at $\beta=1.1$ on the $12\times 6^3$
lattice ({\bf a}) and at $\beta=2$ on the $24\times 12^3$ lattice
({\bf b}).
}
\label{fig:ndpm_2lat_2beta_LG}
\end{figure}

\vfill


%
%
\begin{figure}[pt]
\vspace{1.0cm}
\begin{center}
\vskip -1.5truecm
\leavevmode
\hbox{
\epsfysize=14cm
\epsfxsize=14cm
\epsfbox{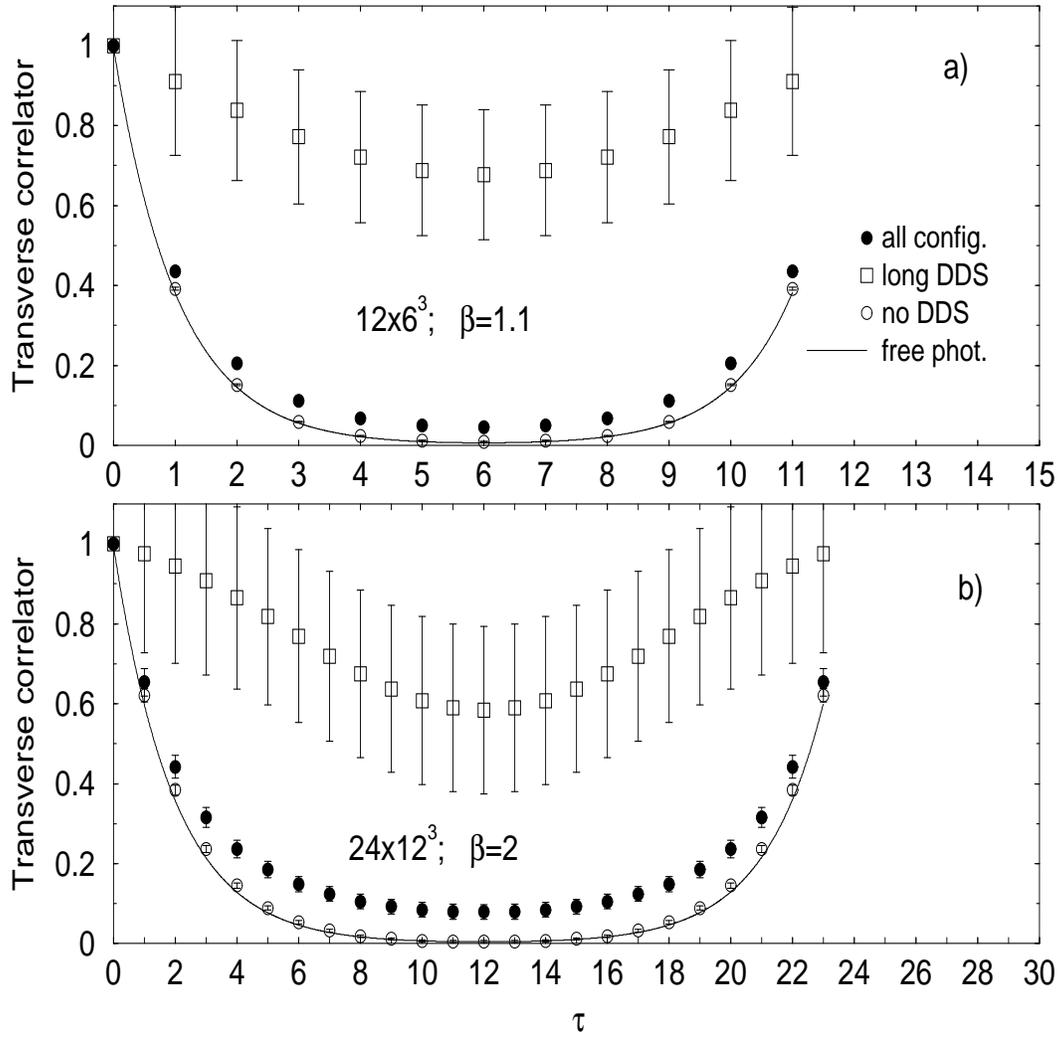}
}
\end{center}
\caption{$\Gamma_T(\tau;\vp )$ at $\beta=1.1$ on the $12\times 6^3$ lattice
({\bf a}) and
at $\beta=2$ on the $24\times 12^3$ lattice ({\bf b}).
}
\label{fig:pcr_2lat_2beta_LG}
\end{figure}

\vfill


%
%
%
\begin{figure}[pt]
\begin{center}
\vskip -1.5truecm
\leavevmode
\hbox{
\epsfysize=14cm
\epsfxsize=14cm
\epsfbox{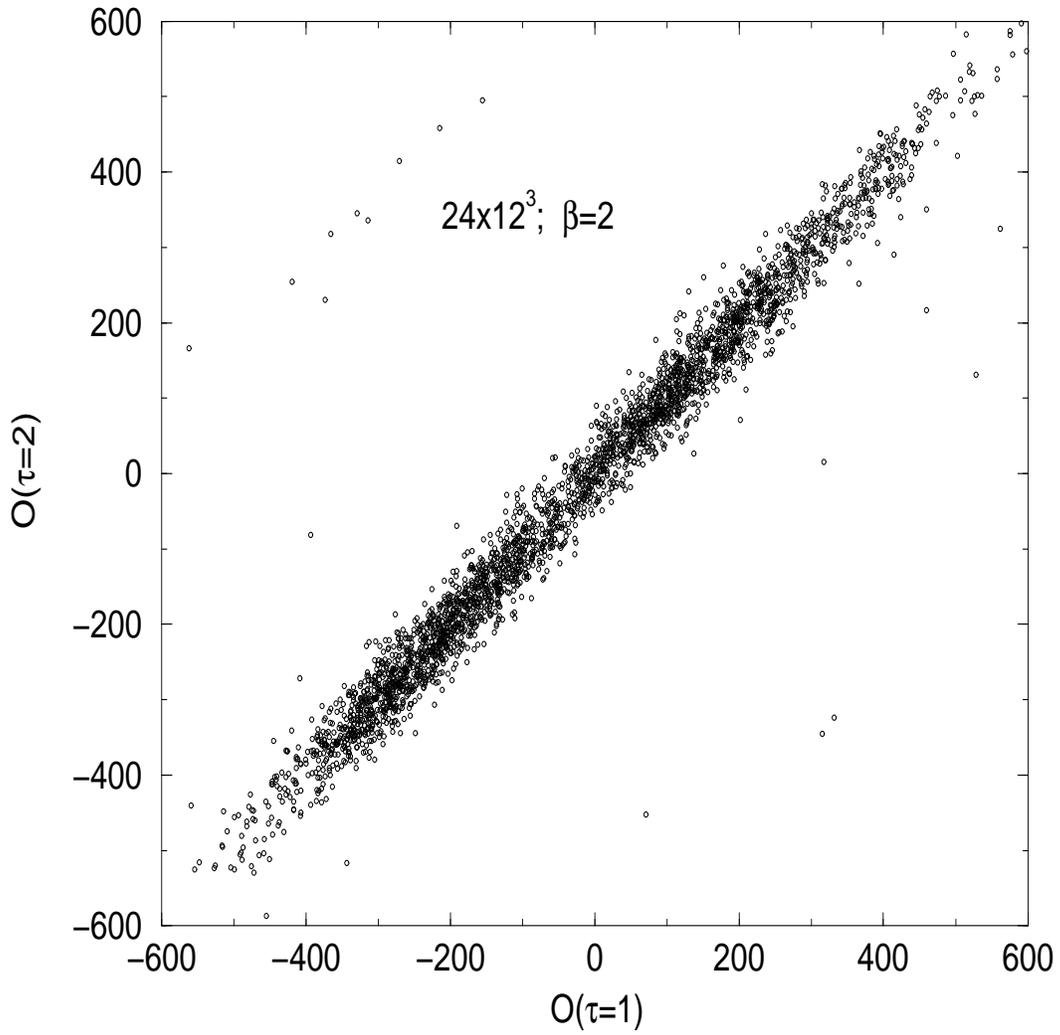}
}
\end{center}
\caption{Scatter--plot for operators $\cao(\tau =1)$
and $\cao(\tau =2)$ at $\beta=2$ on the $24\times 12^3$ lattice.
}
\label{fig:op12_24x12_b02p00}
\end{figure}

\vfill


%
%
%
\begin{figure}[pt]
\begin{center}
\vskip -1.5truecm
\leavevmode
\hbox{
\epsfxsize=14cm
\epsfbox{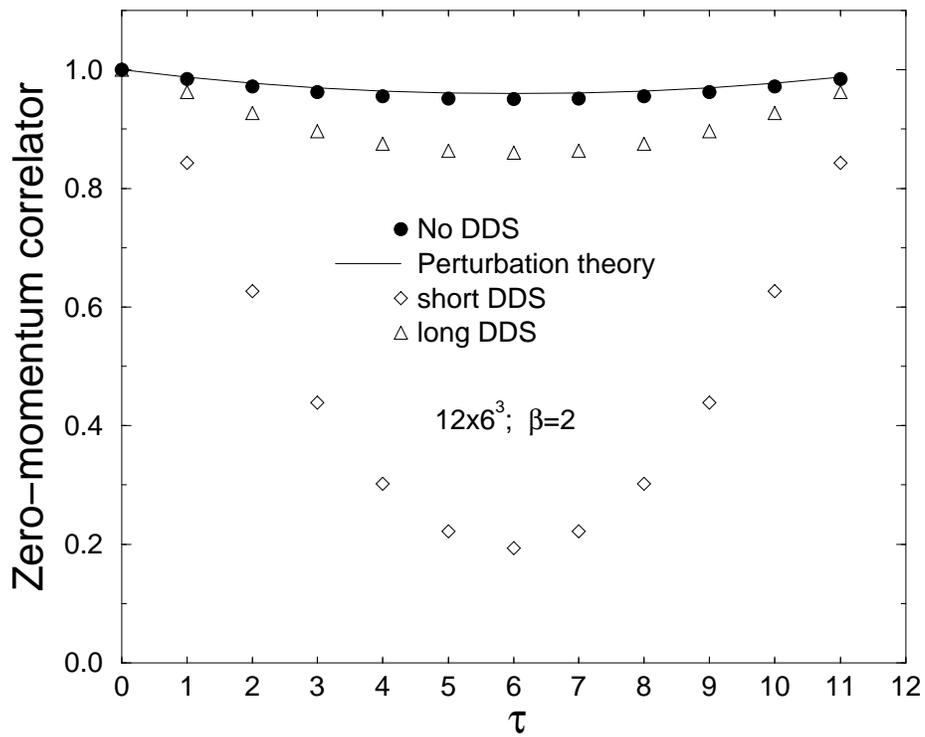}
}
\end{center}
\vskip -1.5truecm
\caption{$\Gamma(\tau)$ at $\beta=2$ on the $12\times6^3$
lattice.
}
\label{fig:pcr_mom0_12x06_b02p00_LG}
\vskip -0.5truecm
\end{figure}

\vfill

\end{document}